\documentclass[prc,twocolumn]{revtex4}
\def\be{\begin{eqnarray}}
\def\ee{\end{eqnarray}}

\usepackage{latexsym,exscale}
\usepackage{psfig}
\usepackage{bm}
\usepackage{amsmath,amssymb,amsfonts}
\newcommand{\ep}{\varepsilon}
\newcommand{\bp}{\bm p}
\newcommand{\bq}{\bm q}
\newcommand{\bk}{\bm k}
\newcommand{\bn}{\bm n}
\newcommand{\bom}{\bar\omega}
\newcommand{\sigmavec}{{\bm \sigma}}
\newcommand{\nablavec}{{\bm \nabla}}
\newcommand{\phivec}{{\bm \phi}}
\newcommand{\tauvec}{{\bm \tau}}
\newcommand{\Fd}{F^{\dagger}}
\newcommand{\Fdl}{F^{\dagger <}}
\newcommand{\Fdg}{F^{\dagger >}}

\begin{document} 
\title{Off-shell pairing correlations from meson-exchange theory
        of nuclear forces}
\author{Armen Sedrakian}
              
\affiliation{ Institut f\"ur Theoretische Physik, 
              Universit\"at T\"ubingen, 
              D-72076 T\"ubingen, Germany}

\date{\today}
\begin{abstract}
We develop a model of off-mass-shell pairing correlations
in nuclear systems, which is based on the meson-exchange picture 
of nuclear interactions. The temporal retardations in the model
are generated by the Fock exchange diagrams. The kernel of the 
complex gap equation for baryons is related to the in-medium 
spectral function of mesons, which is evaluated 
nonperturbatively in the random phase approximation.
The model is applied to the low-density neutron matter in
neutron star crusts by separating the interaction into  a 
long-range one-pion-exchange component and a short-range 
component parametrized in terms of Landau Fermi liquid parameters. 
The resulting Eliashberg-type coupled nonlinear integral equations are solved
by an iterative procedure.
We find that the self-energies extend to off-shell energies of the
order of several tens of MeV. At low energies the damping of the 
neutron pair correlations due to the coupling to the pionic modes 
is small, but becomes increasingly important as the energy is 
increased. We discuss an improved quasiclassical approximation 
under which the numerical solutions are obtained.  
\end{abstract}

%\pacs{12.38-t,26.60.+c,74.20.-z,97.60.Jd,03.75.Fi}

\maketitle

\section{Introduction}

  The lowest order in the interaction mean-field treatments of the nuclear 
  pairing in bulk nuclear matter and in finite nuclei are 
  conceptually simple and have been reasonably consistent with 
  the phenomenology of these systems. The Hartree-Fock-Bogolyubov
  (HFB) theory with  zero- or finite-range (e.g., the Skyrme and the
  Gogny) phenomenological forces has been the standard tool in the
  studies of the nuclear structure to describe the
  bulk of the experimental data  \cite{RING_SCHUCK}. 
  At the same time, the calculations of the pairing gaps in bulk
  nuclear matter
  based on the Bardeen-Cooper-Schrieffer (BCS) theory with interactions 
  modeled by the bare (realistic) nucleon-nucleon forces and
  quasiparticle spectra in the effective mass approximation are 
  not inconsistent with the cooling simulations of neutron stars
  or the phenomenology of their rotation dynamics. 
  Although there is no direct evidence for higher
  order correlations in the phenomenology of nuclear systems, these
  cannot be discarded on the basis of  purely phenomenological arguments.
  The effective forces used in the HFB calculations are fitted 
  to the pairing properties of (some) nuclei, consequently the
  contributions beyond mean field are built-in automatically 
  (e.g.,  the effective Skyrme or Gogny forces include the
  contributions from the ladder  diagrams). The values of 
  the pairing gaps deduced from the phenomenology of 
  neutron stars depend sensitively on the physical input, such as the 
  equation of state, the matter composition, and its dynamical
  properties. Provided the uncertainties in these ingredients the 
  phenomenological values of the gap can  be off by an order of
  magnitude. From the theory point of view the higher order
  correlations might not be manifest due to large cancellation between
  the vertex and propagator renormalization.

  The on-mass-shell properties of the pairing gaps are well understood
  at the mean-field level. Upon ignoring the renormalization effects, the
  on-mass-shell gaps can be directly computed from the nucleon-nucleon
  scattering phase
  shifts. Since the pairing is a low-temperature phenomenon, a
  well-defined Fermi sea implies that the on-shell self-energies 
  of weakly excited quasiparticles can be
  expanded around their Fermi momentum. The magnitude of 
  the gap is then controlled by the renormalization of density of 
  states at the Fermi surface. Since the ratio of the momentum-dependent
  effective mass to the bare mass is less than unity the reduction of the density of
  states causes a reduction of the pairing gap.

  The early work to construct renormalization schemes for 
  strongly interacting fermions introduced
  wave-function renormalization factors in the kernel of the gap equation 
  and modeled the pairing interaction 
  in terms of particle-hole irreducible vertices \cite{MIGDAL}. 
  These were related to the phenomenological Landau Fermi liquid parameters
  which can (in principle) be fixed by comparison with 
  experiments. The Landau parameters can  be obtained from 
  {\it ab initio}  microscopic calculations as well, which are needed, 
  for example, to 
  describe isospin asymmetric nuclear systems that are not directly 
  accessible in laboratories  
  \cite{Backman:sx,Dickhoff:qr,Dickhoff:qr2,Dickhoff:qr3}.
  Various aspects of the pairing in the infinite nuclear 
  matter beyond the mean-field (BCS) approximation were addressed
  during the past two decades by employing either more advanced schemes 
  to calculated the self-energy and vertex corrections and/or modern
  (phase-shift equivalent) interactions. 
  This work includes the microscopic calculations of the effective 
  interactions and  Landau parameters using variants of the 
  Babu-Brown \cite{BABU_BROWN} or related (e.g., polarization
  potential) schemes of summing the polarization graphs in 
  neutron 
  matter \cite{WAMBACH1,WAMBACH2,SCHULZE}. The polarization effects were
  found to suppress the on-shell pairing gaps by large factors.
  Earlier, similar reduction was found from 
  the correlated basis variational calculations based on the 
  Jastrow ansatz for fermionic wave functions
  \cite{CLARK1,CLARK2,CLARK3}. More recent work concentrated on 
  the calculations of 
  the  wave-function renormalizations and their implementations 
  in gap equations employing self-energies 
  derived within Brueckner theory at first 
  \cite{BALDO} and second  \cite{LOMBARDO} order in the 
  hole-line expansion. While again a suppression was found due to the 
  wave function renormalizations, supplementing these self-energy
  corrections with appropriate vertex corrections 
  largely compensates for the latter effect \cite{SHEN}. 
  The finite temperature $T$ matrix was utilized to deduce the pairing
  correlations that include full spectral information on the system 
  by using the  Thouless criterion \cite{ROEPKE}, wave-function 
  renormalizations in the gap equations, or the $T$ matrix in the 
  superfluid phase \cite{BOZEK}. The reduction of the critical
  temperature of superfluid phase transition found in these approaches
  is due to dressing of propagators in the particle-particle channel, 
  i.e., the analog effect of the self-energy renormalizations in the
  Brueckner scheme \cite{BALDO,LOMBARDO}. (It is unrelated to the
  reduction mentioned above caused by the polarization of the interaction 
  that follows from the resummations in the particle-hole channels.) 
  Renormalization group approach has been used to resume the
  particle-hole channels and deduce the pairing gap from the matrix
  elements defined on the Fermi surface \cite{RENORM1}; the overall 
  magnitude of the reduction of the gap is consistent with the 
  results mentioned above. The pairing gaps computed from the 
  renormalization group motivated interactions are in good agreement with the 
  calculations based on the original bare interactions \cite{RENORM2} 
  or the effective Gogny interactions \cite{RENORM3} below the
  renormalization scale.

  The off-mass-shell physics of the nuclear pairing is much less 
  understood. Contributions to the (complex) pairing gap can 
  be generated by time nonlocal interactions. The simplest example is the
  fermionic Fock-exchange diagram in a boson-exchange interaction model.
  The generic theory of the boson-exchange superconductors is known in
  the form first developed by Eliashberg \cite{ELIASHBERG} to describe the
  superconductivity in materials where the electron-phonon
  coupling is of the order of unity (see also the general references 
  \cite{SCHRIEFFER,SCALAPINO,MAHAN}). Because the theory contains
  another small parameter - the ratio of the phonon Debye
  energy to the Fermi energy of electrons - the
  vertex corrections (according to the Migdal theorem) can be neglected. For
  the same reason, the quasiclassical approximation can be used where the
  on-shell energy of the original Green's functions are integrated
  out. In our context the coupling of the pionic modes to baryons is
  not small and the characteristic energies of the off-shell excitations are 
  not small compared to the Fermi energy. Below, we shall partially
  overcome these difficulties, first, by including a vertex renormalization 
  by the short-range correlations in the pion-nucleon vertex and, second,
  by applying an improved version of the quasiclassical approximation.

  Pairing in nuclear systems, induced by retarded interactions, was
  discussed previously in Refs. \cite{SEDRAKIAN,TERASAKI,KAMERDZHIEV}.  
  In the context of laboratory nuclei the retarded pairing interaction 
  is driven by an exchange of virtual phonons - low-lying
  collective bosonic excitation \cite{TERASAKI,KAMERDZHIEV}. 
  In the astrophysical context of neutron star crusts the attractive
  pairing interaction is mediated by an exchange of {\it real} phonons of 
  the nuclear Coulomb lattice of the crusts \cite{SEDRAKIAN}.  The
  approach taken here differs from the models above in the
  nature of the bosonic modes that mediate the pairing interactions. 
  While the density fluctuations  are responsible for the pairing
  interaction in the models above, we concentrate here 
  at the modes that have the quantum numbers of pions.

  In what follows, we construct a model of pairing in neutron matter
  that is based on the meson-exchange picture of the nuclear interactions. 
  Specific to our model is the explicit treatment of the mesonic
  degrees of freedom including retardation. Unlike the potential
  models, the interaction incorporates medium modifications of the
  dispersion relation of the mesons (pions) and  they are treated 
  dynamically, i.e., only in the static limit the retarded interaction 
  reduces to a meson (pion)-exchange potential. We note that the
  meson-exchange picture of pairing is known from applications of the 
  Walecka mean-field model to the pairing problem \cite{RMF1,RMF2,RMF3}.
  These approaches, however, do not include the pion dynamics 
  or retardation.  
  In computing the renormalization of the pionic propagator  and the
  pion-nucleon vertex, we shall use  a simplified version of the  
  resummation of the particle-hole diagrams by modeling the 
  residual interaction in terms of a contact approximation to the  
  $g'$  Landau parameter. This corresponds to the random phase
  approximation (RPA) renormalizations 
  of pion propagators in Migdal's theory of finite Fermi systems 
  \cite{MIGDAL2}.
  The retardation effects are included by evaluating 
  the Fock (exchange) self-energies of neutrons which are coupled to 
  dynamical (off-mass-shell) pions. We determine numerically
  the resulting off-mass-shell gap function and wave-function
  renormalization self-consistently by solving the
  coupled integral equations for these two complex functions. 

To elucidate our model, 
we start with the $\pi_0$-exchange interaction among
neutrons in a low -density neutron matter relevant for the physics of neutron
star crusts. In the density regime $(0.002 - 0.5)~ \rho_0$, 
where $\rho_0 = 0.17$ fm$^{-3}$ is the nuclear saturation density, the matter in
neutron stars is an admixture of unbound neutron liquid  and a 
Coulomb lattice of nuclei with a charge-neutralizing background 
of relativistic electrons \cite{BBP,NV,HAENSEL}. We shall assume that
the neutrons form a homogenous Fermi liquid which is characterized by 
an isotropic Fermi surface. 

The remainder of the paper is organized as follows.
Section. II is devoted to the formulation of the model 
in the framework of the real time Green's 
functions formalism.  Section. III describes a method of 
solution of the coupled, nonlinear integral Eliashberg equations. 
The results are discussed for the case under study -- 
the pairing in the low-density neutron matter -- 
where the retarded pairing interaction is mediated by an exchange of 
neutral pions. Section. IV contains our conclusions. In the Appendix we 
discuss the quasiclassical approximation to the propagators 
and its improved version, which is used in our numerical work.

\section{Model}

We shall formulate the model in terms of the real-time
contour-ordered Green's functions~\cite{KELDYSH}. 
The imaginary time Matsubara formalism is well 
suited for our purposes, since we assume the 
system to be in equilibrium,  however,  the numerical solution of
the equations formulated on the real frequency axis are technically 
simpler compared to the solution of the imaginary time equations. The
latter require a  summation of finite number of discrete Matsubara
frequencies with a certain cutoff frequency which is followed by an 
analytical continuation to the real axis; the former are solved 
directly on the real axis.

\begin{widetext}
The real-time treatment of a superfluid system requires a 
$4\times 4$ matrix of propagators and self-energies in general \cite{KELDYSH}. 
A $2\times 2$  structure is needed to account for the anomalous
correlations due to the pairing. The elements of the Nambu-Gor'kov matrix
propagator $\underline G_{\alpha\beta}(x,x')$ are defined as 
\be 
\left( \begin{array}{cc}
G_{\alpha\beta}(x,x') & F_{\alpha\beta}(x,x')\\
-F^{\dagger}_{\alpha\beta}(x,x') & \overline G_{\alpha\beta}(x,x')\\
\end{array}
\right) = 
\left( \begin{array}{cc}
-i\langle T\psi_{\alpha}(x)\psi_{\beta}^{\dagger}(x')\rangle 
&\langle \psi_{\alpha}(x)\psi_{\beta}(x')\rangle \\
\langle \psi_{\alpha}^{\dagger}(x)\psi_{\beta}^{\dagger}(x')\rangle
&-i\langle \tilde T\psi_{\alpha}(x)\psi_{\beta}^{\dagger}(x')\rangle\\
\end{array}
\right),
\ee
where  $\psi_{\alpha}(x)$ are the baryon field operators, $x$ is the
space-time coordinate,  $\alpha$ and $\beta$ stand for the internal (discrete)
degrees of freedom, $T$ and $\tilde T$ are the symbols for time
ordering and inverse time ordering of operators, respectively. 
Each of the propagators above  can be viewed as a $2\times 2$ matrix 
in the Keldysh space, if we map the time structure of the propagators 
on a  real-time contour. For example, the elements of the normal propagator
in the Keldysh space, ${\underline{\hat G}}_{\alpha\beta}(x,x')$
(the hat hereafter indicates that the propagator is contour ordered)
are defined as 
\be\label{KELDYSH}
\left( \begin{array}{cc}
\underline{G}^c_{\alpha\beta}(x,x') & \underline{G}^<_{\alpha\beta}(x,x')\\
\underline{G}^>_{\alpha\beta}(x,x') & \underline{G}^a_{\alpha\beta}(x,x')\\
\end{array}
\right) = \left( \begin{array}{cc}
-i\langle T \psi_{\alpha}(x)\psi_{\beta}^{\dagger}(x')\rangle 
&i\langle \psi_{\alpha}^{\dagger}(x')\psi_{\beta}(x) \rangle \\
-i\langle \psi_{\alpha}(x)\psi_{\beta}^{\dagger}(x')\rangle
&-i\langle  \tilde T \psi_{\alpha}(x)\psi_{\beta}^{\dagger}(x')
\rangle \\
\end{array}
\right).
\ee 
The off-diagonal elements of the matrix
(\ref{KELDYSH}) have their time arguments on the 
opposite branches of the time contour, while the 
upper (lower) diagonal elements have their both time arguments 
on the positive (negative) branch of the contour (hence the
time-ordering symbols; see for details Ref. \cite{KELDYSH}).
The remainder propagators are defined in a similar way.  
We note that even in the non-equilibrium 
theory only a limited subset of the elements of the matrices above 
carry new information. The Keldysh structure will be used below as a
convenient tool for writing down the finite temperature expressions
for diagrams. Eventually, we will find that the final expressions can
be written only in terms of the normal and anomalous  
retarded propagators.

The $4\times 4$ matrix Green's function satisfies the familiar Dyson equation
\be\label{DYSON}
\underline{\hat G}_{\alpha\beta}(x,x') = 
\underline{\hat G}^0_{\alpha\beta}(x,x') 
+ \sum_{\gamma , \delta}\int\!\!d^4x'' d^4x'''\underline{\hat G}^0_{\alpha\gamma}(x,x''')
\underline{\hat\Sigma}_{\gamma\delta}(x''',x'') 
\underline{\hat G}_{\delta\beta} (x'',x'),
\ee
where the free-propagator matrices are diagonal in both spaces  
and the matrix structure of the self-energies is
identical to that of the propagators (see for details Ref. \cite{KELDYSH}). 
Fourier transforming Eq. (\ref{DYSON}) with
respect to the relative coordinate $x = x-x'$  one obtains the Dyson
equation in the momentum representation; in this representation 
the components of the Nambu-Gor'kov matrix obey the 
following coupled Dyson equations 
\be\label{1}
\hat G_{\alpha\beta}(p) &=& \hat G_{0\alpha\beta}(p) + \hat
G_{0\alpha\gamma}(p) \left[\hat \Sigma_{\gamma\delta}(p) \hat G_{\delta\beta}(p)
+\hat \Delta_{\gamma\delta}(p)\hat\Fd_{\delta\beta}(p) \right],\\
\label{2}
\hat\Fd_{\alpha\beta} (p) &=& \hat G_{0\alpha\gamma}(-p)\left[\hat 
\Delta^{\dagger}_{\gamma\delta}(p) \hat G_{\delta\beta}(p) 
+\hat \Sigma_{\gamma\delta}(-p)\hat \Fd_{\delta\beta} (p) \right],\\
\label{2bis}
\hat F_{\alpha\beta} (p) &=& \hat G_{0\alpha\gamma}(p)\left[\hat 
\Delta_{\gamma\delta}(p) \hat G_{\delta\beta}(-p) 
+\hat \Sigma_{\gamma\delta}(p)\hat F_{\delta\beta} (p) \right],\\
\label{1bis}
\hat G_{\alpha\beta}(-p) &=& \hat G_{0\alpha\beta}(-p) + \hat
G_{0\alpha\gamma}(-p) \left[\hat \Sigma_{\gamma\delta}(-p) \hat G_{\delta\beta}(-p)
+\hat \Delta_{\gamma\delta}^{\dagger}(p)\hat F_{\delta\beta}(p) \right],
\ee
where $p$ is the four momentum,  $\hat G_{\alpha\beta}(p)$ 
and $\hat G_{0\alpha\beta}(p)$ are the full and free normal propagators, 
$\hat \Fd_{\alpha\beta} (p)$ and  $\hat F_{\alpha\beta} (p)$
are the anomalous propagators, and 
$\hat \Sigma_{\alpha\beta}(p)$ and $\hat \Delta_{\alpha\beta}(p)$  
are the normal and anomalous self-energies;  
the greek subscripts are the spin/isospin indices; 
summation over repeated indices is understood. 
We shall assume that the interactions conserve spin and isospin, i.e.,
$\hat G_{\alpha\beta}(p) = \delta_{\alpha\beta}\hat G(p)$ 
and $\hat\Sigma_{\alpha\beta} (p)= \delta_{\alpha\beta}\hat\Sigma(p)$
and concentrate below on the pairing in the state of total spin 
$S = 0$, total isospin $I =1$, and orbital angular momentum $L=0$. 
Thus the anomalous propagators and self-energies must be antisymmetric
with respect to the spin indices and symmetric in the isospin indices
\be
 \hat \Fd_{\alpha\beta} (p) &=& g_{\alpha\beta}  \hat \Fd (p), \quad 
 \hat F_{\alpha\beta} (p) = g_{\alpha\beta} \hat F (p),\\
\hat\Delta^{\dagger}_{\alpha\beta} (p) &=& g_{\alpha\beta}\hat\Delta^{\dagger} (p), \quad 
\hat\Delta_{\alpha\beta} (p) = g_{\alpha\beta}\hat\Delta (p),
\ee
where $g_{\alpha\beta} \equiv i\sigma_y$ is the spin matrix
with  $\sigma_y$ being the second
component of the vector Pauli matrices; the unit matrix in the isospin
space is suppressed. The Dyson equations (\ref{1})-(\ref{1bis}) can be
written in terms of  auxiliary Green's functions, which describe the
unpaired state of the system 
\be \label{N1}
\hat G^N_{\alpha\beta}(p) &=& \hat G_{0\alpha\beta}(p)+\hat
G^N_{\alpha\gamma}(p)\hat\Sigma_{\gamma\delta}(p) \hat
G^N_{0\delta\beta}(p),\\
\label{N2}
\hat G^N_{\alpha\beta}(-p) &=& \hat G_{0\alpha\beta}(-p)+\hat
G^N_{\alpha\gamma}(-p)\hat\Sigma_{\gamma\delta}(-p) \hat
G^N_{0\delta\beta}(-p).
\ee
Combining Eqs. (\ref{1})-(\ref{1bis}) and (\ref{N1}) and (\ref{N2}) we
find an equivalent form of the Dyson equations
\be\label{D1}
\hat G_{\alpha\beta}(p) &=& \hat G^N_{\alpha\gamma}(p)
\left[\delta_{\gamma\beta}+\hat\Delta_{\gamma\delta}(p)
\hat\Fd_{\delta\beta}(p)\right],\\
\label{D2}
\hat\Fd_{\alpha\beta}(p) &=& \hat G^N_{\alpha\gamma}(-p) 
\hat\Delta^{\dagger}_{\gamma\delta}(p) \hat G_{\delta\beta}(p),\\
\label{D3}
\hat F_{\alpha\beta}(p) &=& \hat G^N_{\alpha\gamma}(p) 
\hat\Delta_{\gamma\delta}(p) \hat G_{\delta\beta}(-p),\\
\label{D4}
\hat G_{\alpha\beta}(-p) &=& \hat G^N_{\alpha\gamma}(-p)
\left[\delta_{\gamma\beta}+\hat\Delta_{\gamma\delta}^{\dagger}(p)
\hat F_{\delta\beta}(p)\right].
\ee
\end{widetext}
The time-reversal symmetry implies that 
$
\Delta_{\alpha\beta}(p)=[ \Delta^{\dagger}_{\alpha\beta}(p)]^* ,
$
i.e. the pairing correlations in the system are fully specified if we 
know one of the two anomalous Green's functions; in other words,
it is sufficient to solve Eq. (\ref{D1}) with either
Eq. (\ref{D2}) or Eq. (\ref{D3}). If we are interested in the equilibrium
properties of the system, it is convenient to solve 
Eqs. (\ref{D1})-(\ref{D4}) for the 
retarded propagators. The form of these equations remains the same 
as we replace the contour-ordered functions by the retarded (more
precisely, the contour ordered Dyson equation is invariant under
unitary transformations which bring the Keldysh matrix to
the so-called `triangular' form in which the diagonal elements are the
retarded and advanced propagators \cite{KELDYSH}; the
simplest example of such a transformation is given by the 
unitary matrix $U = (1+i\sigma_y)/\sqrt{2}$ which acts in the
Keldysh space). We first substitute in Eqs. (\ref{N1}) and (\ref{N2}) 
for the normal state retarded propagators 
$G^{N, R}_{\alpha\beta}(\pm p)=\delta_{\alpha\beta} 
[\pm(\omega +i\eta)-\xi_p -\Sigma^R(\pm p)]^{-1}$, where $\xi_p$ is the
free quasiparticle spectrum in the normal state. After  decomposing 
the retarded self-energy into its even and odd in $\omega$ components,
$\Sigma^R(p) = \Sigma_S^R(p) + \Sigma_A^R(p)$,
we further define the wave-function renormalization $Z(p) = 1-
\omega^{-1}\Sigma_A^R(p)$ and the  renormalized quasiparticle 
spectrum $\xi_p^* = \xi_p+\Sigma_S^R(\bp ,\xi_p^*)$.
With these definitions, 
the solution of algebraic equations (\ref{D1}) and (\ref{D3}) is 
\be\label{9a}
G^R(p) = \frac{\omega Z(p)+\xi^*_p}{(\omega+i\eta)^2 Z(p)^2-\xi_p^{*
2}-\Delta^{R}(p)^2},\\
\label{9b}
F^R(p) =- \frac{\Delta^{ R}(p)}
{(\omega+i\eta)^2 Z(p)^2-\xi_p^{* 2}-\Delta^{R}(p)^2},
\ee
where we used the relation $\Delta_{\alpha\beta}\Delta^{\dagger}_{\beta\alpha}
= - \Delta^2$. 
For the purpose of analytical representation of diagrams for the 
self-energies it is convenient to introduce additional real-time 
Green's functions  by the relations 
\be
\label{10}
G^{>,<}(p)&=& iA_G(p)f^{>,<}(\omega), \\ 
\label{11}
F^{>,<}(p)&=& iA_F(p) f^{>,<}(\omega),
\ee
where $A_G(\omega)$ and $A_F(\omega)$ are the normal and anomalous
spectral functions and 
$f^{>,<}(\omega)$ are the Wigner distribution functions.
In nonequilibrium theory  these propagators are the off-diagonal elements of the
Keldysh matrix (\ref{KELDYSH})
(i.e., have their time arguments on the opposite branches of the 
timecontour) and are the solutions of  transport equations. However,
in equilibrium the relations (\ref{10}) and (\ref{11}) do not contain
additional information,  for the spectral functions
$A_G(\omega)$ and $A_F(\omega)$ can be related to the 
retarded propagators by the their spectral representation
(here we drop the three-momentum dependence of the functions)
\be\label{spec1}
G^R(\omega) &=& \int_{-\infty}^{\infty}\frac{d\ep}{2\pi}
\frac{A_G(\ep)}{\omega-\ep+i\eta},\\
\label{spec2}
F^R(\omega) &=& \int_{-\infty}^{\infty}\frac{d\ep}{2\pi}
\frac{A_F(\ep)}{\omega-\ep+i\eta},
\ee
and the Wigner distribution functions reduce to the Fermi distribution
functions for particles and holes: 
$f^{<}(\omega)\equiv f(\omega) = [{\rm exp}(\beta\omega)+1]^{-1}$ and 
$f^{>}(\omega) = f(\omega)-1$, where $\beta$ is the inverse temperature. In other
words, the  relations (\ref{10}) and (\ref{11}) establish a one to one 
correspondence between the retarded and  $G^{ >,<}(p)$,
$F^{>,<}(p)$ propagators.

To model the strong interaction, we separate its long-range and short-range
components. The long-range part of nuclear interaction is dominated by
the pion dynamics with the characteristic scale set by  the pion
Compton wavelength $m_{\pi}^{-1} =1.4$ fm. The nonrelativistic form of 
the effective pion-nucleon interaction Hamiltonian is 
\be 
H_{\pi NN} = -\frac{f_{\pi}}{m_{\pi}} 
(\sigmavec \cdot \nablavec ) (\tauvec \cdot \phivec ),
\ee
where $ \phivec$ is the pseudoscalar isovector pion field satisfying
the Klein-Gordon equation, $f_{\pi}$ is the pion-nucleon coupling 
constant, $m_{\pi}$ is the pion mass, $\sigmavec$ and $\tauvec$  are 
the vectors of Pauli matrices in the spin and isospin spaces.
For {\it static pions} the one-pion-exchange 
interaction among neutrons in the momentum space has the form
\be\label{ope} 
V_{\pi}(\bq) = -\frac{f_{\pi}^2}{3m_{\pi}^2}\, 
\frac{\bq^2}{\bq^2+m_{\pi}^2}
\left[\sigmavec _1 \cdot \sigmavec _2 +S_{12}(\bn)\right]
\tauvec_1 \cdot \tauvec_2,
\ee
where $\bq$ is the momentum transfer, 
$\bn =\bq /q$, and $S_{12}(\bn)$  is the tensor operator.
It is known to reproduce the low-energy phase shift and, to a large
extent, the deuteron properties \cite{WEISE}. Below, the pairing
correlations are evaluated from the diagrams which contain {\it
dynamical pions} with full account of the frequency dependence of the 
pion propagators; the static results
could be recovered, and a relation to the phase shifts established, 
only in the limit $\omega \to 0$ in the pion propagators.
The intermediate and short-range dynamics is dominated, 
respectively, by the correlated two-pion
exchange $\rho$ and other heavy meson-exchanges. Short-range
correlations are crucial for a realistic description of low-energy
phenomena, since the response functions calculated from 
one-pion-exchange alone would lead to an instability of nuclear matter
at the nuclear saturation density (pion condensation) which is not observed.
The short-range correlations can be modeled by an effective
interaction of the form \cite{WEISE} 
\be\label{vshort}
V_{\rm corr}(\omega , q)&=& \frac{f_{\pi}^2}{m_{\pi}^2}
 \Bigl[g'(\omega , q) \sigmavec _1\cdot \sigmavec _2\nonumber\\
&&\hspace{0.1cm} + 
h'(\omega , q) \sigmavec _1\cdot \sigmavec _2 ~ S_{12}(\bn) \Bigr]
\tauvec_1 \cdot \tauvec_2,
\ee
where $g'$ and $h'$ are the dimensionless Landau parameters.  
Equation (\ref{vshort})contains only the spin-isospin part of the 
short-range interaction, which is 
relevant for the RPA renormalization of the vertices and the polarization
tensor. The parameters $g'$ and $h'$ describe
short-range correlations on the scale $l_{\rm corr}$ which is much smaller
than the Compton length of the pion, hence, they should be smooth
functions of the momentum transfer $q$ and the energy of excitations 
$\omega$. Frequently, they are approximated as  $g'=$ const
and  $h'=(l_{\rm corr} k)^2\times $ const. Furthermore, often
only the contribution from the $g'$ parameter is kept, since the contribution
of the tensor component $\sim h'$ is numerically small. The
phenomenological features of the pion-nucleon systems described above
are consistent with the  microscopic theories
\cite{Backman:sx,Dickhoff:qr}.
The full meson propagator obeys the Dyson equation 
\be \label{5}
\hat D(p) &=& \hat D_0(p) + \hat D_0(p)\hat \Pi(p) \hat D(p),
\ee
where $\hat D_{0}(q)$  is the free meson propagator and the 
polarization tensor is defined  as
\be \label{6}
\hat\Pi (q) &=& -{\rm Tr}\int\!\!
\frac{d^4p}{(2\pi)^4}\, i\, \hat\Gamma_0 (q) \hat
G(p+q) \hat G(-p)\hat\Gamma(q),
\ee
where $\hat\Gamma_0(q)$ and $\hat\Gamma(q)$ are the bare and full
pion-nucleon vertices; $\hat\Gamma_0(q) =-\left(f_{\pi}/m_{\pi}\right)~ 
(\sigmavec \cdot \bq) $. The full  pion-nucleon vertex is
defined by the integral equation 
\be\label{vertex}
\hat\Gamma(q) = \hat\Gamma_0(q)+ 
{\rm Tr}\int\!\! 
\frac{d^4p}{(2\pi)^4}\, i\, \hat \Gamma_{1} (q) \hat
G(p+q) \hat G(-p)\hat\Gamma(q),\nonumber\\
\ee
where $\hat\Gamma_{1} (q)$ is the short-range part of the particle-hole
interaction, which can be approximated by Eq. (\ref{vshort}).
In general the propagators in Eq. (\ref{6}) and (\ref{vertex}) are $2\times 2$
matrices in the Nambu-Gor'kov space. The
contributions from the anomalous sector, however, involve 
propagator products  which contain powers of $F$,
each contributing a suppression factor 
$O (\Delta/\mu)$, where $\mu$ is the Fermi energy.
Therefore, we shall keep below only the contribution from the
nonsuperconducting propagators to the polarization tensor, 
i.e., the renormalization of the pion dispersion relation 
is carried out in the normal state. 
As in the baryon sector, we  introduce the 
pion Green's functions $D^{>,<}(q)$, which are the off-diagonal 
elements of the nonequilibrium matrix propagator 
(i.e., have their time arguments on the opposite branches of the 
time contour) by the relations
\be 
D^<(q)&=& -iB(q) g(\omega) -i B(-q)[1+ g(-\omega)],\\
D^>(q)&=& -iB(q)[1+ g(\omega)] -i B(-q)g(-\omega),
\ee
where $g(\omega) = [{\rm exp}(\beta\omega)-1]^{-1}$ is the Bose 
distribution function and $B(q)$ is the pion spectral function
defined over the frequency range $\omega \in [\,0;~ \infty]$.
The latter is related to  the retarded component of the 
polarization tensor, Eq. (\ref{6}), by the relation
\be 
B(q) = \frac{-2{\rm Im}\Pi^R(q)}
{[\omega^2-\bq^2-m_{\pi}^2-{\rm Re}\Pi^R(q)]^2
+[{\rm Im}\Pi^R(q)]^2}.
\ee
The one-loop-polarization tensor for neutral pions in a 
noninteracting neutron gas at zero temperature is given by 
\be 
\Pi^R_0(\omega , q) = -\frac{2f_{\pi}^2 q^2}{m_{\pi}^2}  \phi_L(\omega , q),
\ee
where $\phi_L(\omega , q)$ is the Lindhard function. 
The RPA renormalization of the polarization tensor then leads to 
\be\label{PI_R}
\Pi^R(\omega , q) = -\frac{2f_{\pi}^2 q^2}{m_{\pi}^2} 
\left[1-\frac{2f_{\pi}^2}{m_{\pi}^2} g'
\phi_L(\omega , q)\right]^{-1}\, \phi_L(\omega , q).\nonumber\\
\ee
The form of the polarization tensor thus completely determines the 
spectral function of pions, which is the central quantity for the 
calculations of self-energies.
The time nonlocal parts of the normal and anomalous self-energies are due 
to the Fock contributions defined as  
\be \label{3}
\hat\Sigma (p) &=& -{\rm Tr}\int \frac{d^4q}{(2\pi)^4} \hat\Gamma_{0}(q) 
\hat D(q) \hat G(p-q)\hat\Gamma(q),\\
\label{4}
\hat\Delta (p) &=& -{\rm Tr}\int \frac{d^4q}{(2\pi)^4} \hat \Gamma_{0} (q) 
\hat D(q)\hat F(p-q) \hat\Gamma(-q) .
\ee
Note that only the central part of pion-exchange interaction contributes to the 
pairing interaction in the $S = 0, \, I = 1$ state. To project out the
pairing interaction for these quantum numbers we use the identity
\be 
(\sigmavec_1\cdot \bq) (\sigmavec_2\cdot \bq) &=&
\frac{1}{3} (\sigmavec_1\cdot \sigmavec_2) q^2\nonumber\\
&&\hspace{-1.3cm}+\frac{1}{3} \left[
3 (\sigmavec_1\cdot \bq)(\sigmavec_2\cdot \bq)
-(\sigmavec_1\cdot \sigmavec_2) q^2
\right],
\ee
drop the second (tensor) term and substitute
$
(\sigmavec_1~\cdot~\sigmavec_2) (\tauvec_1~\cdot~\tauvec_2) =-3
$. To obtain the full pion-neutron vertex we use the same procedure as 
for the polarization tensor to include the effects of the short-range 
correlations. However, we do not dress the pion-nucleon vertex by
pion mediated interactions, therefore our model corresponds to the lowest order
contribution in the number of pion-nucleon vertices. 
\begin{figure}[t]
{\psfig{figure=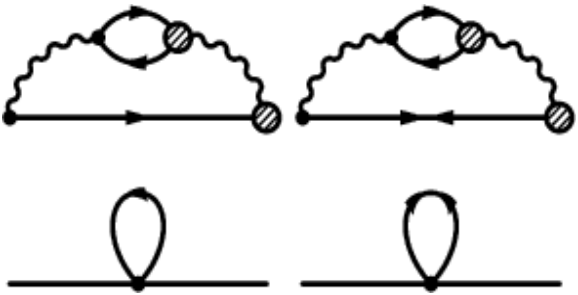,height=1.5in,width=3.in,angle=0}}
\caption{Top panel: The baryon Fock-exchange self-energies 
for normal (left graph) and anomalous (right graph) sectors.
The solid lines correspond to the fermions, the wavy lines 
- to the pions; the  blobs schematically indicate the RPA
renormalized vertices, the dots - the bare pion-nucleon vertices. 
Bottom panel: The  normal and anomalous Hartree diagrams. 
The dots stand for the contact Landau interactions; the straight
lines are shown for clarity.
}
\end{figure}
The top panel in Fig. 1  
shows the Fock diagrams corresponding to Eqs. (\ref{3})
and (\ref{4}); the exchange diagrams, which are related to the direct
ones by crossing symmetry are not shown, but are taken into account in
the definition of the polarization function (\ref{PI_R}). The bottom
panel in Fig. 1 shows the Hartree diagrams which contribute to the mean field
and the BCS interaction via the short-range component of the
interaction (the contribution of the pions to these diagrams
vanishes).
\begin{widetext}
The analytical expressions for the retarded components of the
self-energies can be read off from the diagrams  
\be\label{20a} 
\Sigma^R(\omega) &=& \int_{-\infty}^{\infty}\frac{d\bom}{2\pi}
\frac{\Sigma^>(\bom)-\Sigma^<(\bom)}{\omega-\bom+i\eta},\nonumber\\
&=&\int_{-\infty}^{\infty}\frac{d\bom}{2\pi}\int_{0}^{\infty}\frac{d\omega'}{2\pi}
\frac{D^>(\omega')G^>(\bom-\omega')-D^<(\bom)G^<(\bom-\omega')}{\omega-\bom+i\eta},
\\\label{20b} 
\Delta^R(\omega) &=& \int_{-\infty}^{\infty}\frac{d\bom}{2\pi}
\frac{\Delta^>(\bom)-\Delta^<(\bom)}{\omega-\bom+i\eta}\nonumber\\
&=&\int_{-\infty}^{\infty}\frac{d\bom}{2\pi}\int_{0}^{\infty}\frac{d\omega'}{2\pi}
\frac{D^>(\omega')\Fdg(\bom-\omega')-D^<(\bom)\Fdl(\bom-\omega')}{\omega-\bom+i\eta},
\ee
where we used the dispersion relations for the retarded self-energies $\Sigma^R(\omega)$ 
and $\Delta^R(\omega)$ [in Eqs. (\ref{20a}) and (\ref{20b}) the
momentum variables and the vertices have been suppressed].
Substituting the spectral representations of the Green's functions in
the above expression we arrive at 
\be \label{24}
\Sigma^R(\omega,\bp)  &=&{\rm Tr}\int\!\!\frac{d^3q}{(2\pi)^3} 
\int_{-\infty}^{\infty}\!\!\frac{d\ep}{2\pi}\Gamma_{0}(\bq)
A_G(\ep ,\bp-\bq)C(\omega , \ep , \bq) \Gamma(\bq),\\
\label{25}
\Delta^R(\omega,\bp)  &=&{\rm Tr}\int\!\!\frac{d^3q}{(2\pi)^3} 
\int_{-\infty}^{\infty}\!\!\frac{d\ep}{2\pi}
\Gamma_{0}(\bq)A_F(\ep ,\bp-\bq)
C(\omega ,\ep , \bq)\Gamma(\bq),
\ee
where
\be \label{26}
C(\omega , \ep , \bq) = 
\int_{0}^{\infty}\!\!\frac{d\omega '}{2\pi}B(\omega' , \bq)
\left[
\frac{f(\ep)+g(\omega')}{\ep-\omega'-\omega-i\eta}
+\frac{1-f(\ep)+g(\omega')}{\ep+\omega'-\omega-i\eta}
\right].
\ee
The expressions (\ref{24})-(\ref{26}) for the Fock self-energies  
are general to the extent that the approximations to the 
spectral function of pions and the pion-nucleon vertices 
have not been specified. In the numerical computations below we shall use
the RPA renormalized spectral function of pions and the 
one-pion-exchange vertices which are dressed by the short-range correlations;
clearly, our choice of the resummation scheme is not unique, but it 
is known to include the dominant set of graphs that renormalize the
modes related to the long-range perturbations.

Next we proceed to integrate out the dependence of the self-energies 
on the on-shell energies by implementing the improved 
quasiclassical  approximation (IQCA), which is described in the
Appendix. Since at low temperatures the particle momenta lie close to
their Fermi surface, the quasiclassical approximation (QCA), in its
common form, expands the functions with respect to the small parameter 
$\delta p/p_F$, where $\delta p = \vert p-p_F\vert$ is the deviation of
the particle momenta from their value on the Fermi surface, and keeps the
zeroth order in $\delta p$  terms. Thus, the on-mass-shell physics is constrained to
the Fermi surface because of the degeneracy of the
system. Independently, the QCA assumes that the typical off-shell excitation
energies are much smaller than the Fermi energy, so that the momentum integrals 
over the Green's functions can be evaluated within infinite
limits.  The IQCA avoids the latter approximation since, as we shall
see below, the off-shell excitation energies are of the same order of 
magnitude as the Fermi energy.

Upon implementing the IQCA, Eqs. (\ref{24}) and (\ref{25}) can be written as 
\be\label{28}
\Sigma^Q(p_F,\omega) &=& -\int_0^{\infty}d \omega' K(\omega')
\Bigl\{g(\omega')\left[G^Q(\omega+\omega')+G^Q(\omega-\omega')\right]
\nonumber\\
&&+\int_{-\infty}^{\infty}\frac{d\ep}{\pi} {\rm Im~}[ G^Q(\ep) ]J(\omega , \omega' , \ep)
\Bigr\}, \\
\label{29}
\Delta^Q (p_F,\omega) &=& \int_0^{\infty}d \omega' K(\omega')
\Bigl\{
g(\omega')\left[F^Q(\omega+\omega')+F^Q(\omega-\omega')\right]
\nonumber\\
&&
+\int_{-\infty}^{\infty}
\frac{d\ep}{\pi} {\rm Im~}[ F^Q(\ep) ]J(\omega , \omega' , \ep)
\Bigr\},
\ee
where $\Sigma^Q  (\omega,p_F)$ and $\Delta^Q  (\omega, p_F)$ are the
quasiclassical counterparts of the retarded self-energies,
\be\label{DEF_J}
J(\ep, \omega , \omega') = 
\frac{f(\ep)}{\ep-\omega-\omega'-i\eta}
+\frac{1-f(\ep)}{\ep-\omega+\omega'-i\eta} ,
\ee
and we defined a momentum averaged (real) interaction kernel as
\be\label{K} 
K(\omega) = \frac{m^*}{(2\pi)^3p_F}\int_0^{2p_F}dq~
q\int_0^{2\pi}d\phi~B(\bq,\omega)~{\rm Tr}~ \{\Gamma_{0}(\bq)\Gamma(\bq)\} .
\ee
Formally, Eqs. (\ref{28}) and (\ref{29}) are a coupled set of nonlinear integral 
equations for the complex pairing amplitude and the normal self-energy
(i.e. are equivalent to four integral equations for four real
quantities). This becomes  explicit if we note that the Green's
functions on the right-hand side of  Eqs. (\ref{28}) and (\ref{29}) can be 
expressed in terms of certain integrals
\be
G^Q(\omega) = \omega Z^Q(\omega) I_1(\omega)+I_2(\omega),
\quad F^Q(\omega) = -\Delta^Q I_1(\omega),
\ee
which are defined by Eqs. (\ref{I1}) and (\ref{I2}) of the 
Appendix and are functions of the pairing field $\Delta^Q
(p_F,\omega)$ and the wave-function renormalization  $Z^Q(p_F,\omega)$.
\end{widetext}
Let us briefly comment on the physical 
content of Eqs. (\ref{28}) and (\ref{29}).
They describe the pairing and wave-function renormalization
among degenerate fermions which live on their Fermi surface and
interact via a force that depends on the momentum transfer. The effective
pairing interaction $K(\omega)$ is an average over the momentum 
transfers that are accessible to particles constrained to 
their Fermi surface ($0\le q\le 2p_F$) with a weight given by the 
pion spectral function (i.e., the probability of finding a pion of 
momentum $\bq$ at a fixed energy $\omega$). Although 
the function $K(\omega)$ is independent of
momentum, it is not equivalent to a contact approximation to the
(nonretarded) $S$-wave interaction in the potential models; the latter
would correspond, in the present context,
to the limit of zero momentum transfer. Let us 
recall again that the interaction kernel, via the momentum averaged 
spectral function of pions, incorporates the modification of the  
pionic spectrum by the nuclear medium - a feature that is  
missing in the potential models.
  
We turn now to the anomalous Hartree contribution, shown by the second 
diagram in the lower panel of Fig. 1 (the first diagram renormalizes
the on-shell particle mass). 
Approximating the $S$-wave interaction in terms of the Landau
parametrization,  $F_0(\omega, \bq)+ G_0(\omega, \bq) 
(\sigmavec_1\cdot\sigmavec_2)$, 
one finds for the mean-field BCS contribution 
to the pairing in the spin-zero state
\be 
\Delta^{R}_{BCS}(\omega,\bp) &=& -i\int\frac{d^4p'}{(2\pi)^4} 
[F_0(\omega', \bp')-3G_0(\omega',\bp')]\nonumber\\
&&\times \Fdl
(\omega+\omega',\bp+\bp').
\ee
If the Landau parameters are further approximated by constants, 
the quasiclassical approximation to the Hartree self-energy becomes
\be\label{DBCS} 
\Delta_{BCS}^Q  = 
(F_0-3G_0) \nu(\mu)\int_{-\infty}^{\infty}\frac{d\omega}{\pi}
f(\omega) {\rm Im~}[ F^Q(\omega) ],
\ee
where $\nu(\mu) = p_Fm^*/(2\pi^2)$ is the density of states 
at the Fermi surface.  

In the zero-temperature limit, Eqs. (\ref{28}) and
(\ref{29}) take the simple form
\be \label{30}
Z^Q(\omega) &=& 1-\frac{1}{\omega}
\int_{-\infty}^{\infty}\frac{d\ep}{\pi}  
~{\rm Im}\left[G^Q(\ep)\right] \, L(\omega,\ep),
\\\label{31}
\Delta^Q(\omega) &=& 
\int_{-\infty}^{\infty}\frac{d\ep}{\pi} 
~{\rm Im}\left[F^Q(\ep)\right]\,  L(\omega,\ep),
\ee
\begin{widetext}
where the effective (complex) retarded interaction is defined as 
\be \label{32}
L(\omega ,\ep) = \int_{0}^{\infty} d\omega' K(\omega')
\left[\frac{\theta (\ep)}{\ep-\omega+\omega'-i\eta}
+\frac{\theta (-\ep)}{\ep-\omega-\omega'-i\eta}\right],
\ee
where $\theta(\ep)$ is the Heaviside's step function.  Note that the
thermal occupation probability of the pions vanishes in the 
zero-temperature limit and they contribute only to 
the effective interaction $L(\omega ,\ep)$. Note also that due to the 
time-reversal symmetry of the problem the energy integrations in 
Eqs. (\ref{30}) and (\ref{31})  can be restricted to the positive 
energy domain upon using the 
property $\Delta^Q(-\ep) = \Delta^Q(\ep)$, the time-reversal properties of
the quasiclassical propagators (see the Appendix) and the symmetries of 
the effective interaction $L(\omega ,\ep)$. 
The form of Eq. (\ref{30}) remains the same, but the 
effective interaction is replaced according to $ L(\omega,\ep)\to
L_-(\omega,\ep)$, while Eq. (\ref{31}) becomes 
\be\label{31bis}
\Delta^Q (\omega) &=& 
\int_{0}^{\infty}\frac{d\ep}{\pi}  
~\left\{{\rm Im}\left[F_{\rm odd}^Q(\ep)\right]\,  L_+(\omega,\ep)
+~{\rm Im}\left[F_{\rm even}^Q(\ep)\right]\,  L_-(\omega,\ep)
\right\},
\ee
where the subscripts on the propagators refer to their odd and even 
in the energy variable parts and  the new kernels are defined as 
\be\label{32bis} 
L_{\pm}(\omega, \ep) = \int_0^{\infty} d\omega' K(\omega') 
\left[\frac{1}{\ep+\omega+\omega'+i\eta}
\pm\frac{1}{\ep-\omega+\omega'-i\eta}\right].
\ee
The zero-temperature expression for the  
BCS contribution to the pairing gap becomes 
\be\label{DBCS2} 
\Delta_{BCS}^Q  = 
(F_0-3G_0) \nu(\mu)\int_{0}^{\infty}\frac{d\ep}{\pi}
\left\{ {\rm Im~}[ F_{\rm even}^Q(\ep)] -  {\rm Im~}[ F_{\rm
odd}^Q(\ep) ] \right\}.\nonumber\\
\ee
The net pairing gaps in the quasiparticle spectrum 
is the sum of the Fock-exchange and Hartree 
terms $\Delta(\omega) \equiv \Delta^Q (\omega)
+ \Delta_{BCS}^Q $, defined  by Eqs. (\ref{31bis}) 
and (\ref{DBCS2}).
\end{widetext}
\section{The algorithm and results}

Equations (\ref{30}) and (\ref{31}) form a coupled set of four nonlinear
integral equations for four real functions: these are the real 
and imaginary parts of the gap
function $\Delta_1(\omega)$ and  $\Delta_2(\omega)$, and the 
real and imaginary parts of the wave-function renormalization
$Z_1(\omega)$ and $Z_2(\omega)$. Since the interaction kernel
$L(\omega, \ep)$ is constructed from the spectral function of pions,
renormalized by the nuclear medium in the normal state, it serves as an
input to Eqs.  (\ref{30}) and (\ref{31}) and can be computed prior
to their solution. Thus, as a first step, 
the $L(\omega ,\ep)$ function is generated on a
two-dimensional energy grid. The starting point is the construction of
the pion-spectral function from the real and imaginary parts of 
the RPA renormalized polarization function. The RPA polarization
function is in turn constructed from the free-particle polarization 
function (or the Lindhard function) which has an analytical form both 
at zero and finite temperatures. From the pion spectral function and
the RPA renormalized vertex $\Gamma(q)$ we then construct the
momentum-averaged effective interaction $K(\omega)$ according to
Eq. (\ref{K}) in a typical energy range $0\le \omega \le 200 $ MeV.
The first part of the computation is accomplished by constructing the 
real and imaginary parts of the function $L(\omega ,\ep)$ on a 
two-dimensional energy grid. Typical computations were carried out 
on a $100\times 100$ point mesh. The range of energy integrations  was
extended up to 400 MeV to ensure that the integrands in Eqs. (\ref{30}) and
(\ref{31}) have dropped to 
values below the accuracy of the integration. Note that the
evaluation of Eq. (\ref{32}) requires principle value integrations due
to the simple poles in the integrand and, in addition, evaluation of 
integrals which are singular at the boundary.
 
In the second part of the computation the four integral equations 
represented by Eqs. (\ref{30}) and (\ref{31}) are solved 
self-consistently using as an input the values of the function
$L(\omega ,\ep)$ stored on a two-dimensional energy grid.  
An iterative procedure is used.
The iteration starts by assigning to the functions the
values  $\Delta_1(\omega) =$ const,  $\Delta_2(\omega)=0$,
$Z_1(\omega)=1$ and $Z_2(\omega)=0$, where the constant  
is set equal to the typical scale $\sim 1$ MeV. The newly computed functions are
reinserted in the kernels on the right-hand side of the equations and 
the iterations are repeated until convergence is achieved. Note
that the kernels of  Eqs. (\ref{30}) and (\ref{31}) are singular; in
the first of these equations the singularity is always at the lower 
boundary of the integration region, while in the second equation 
singularities occur both at the lower end point and within the 
integration regions (the latter is understood in the 
Cauchy sense). The positions of the singularities, of course, change
from one iteration to another. An efficient use of the quadratures can 
be achieved by dividing the integration region  into subintervals: 
a 24 point Clenshaw-Curtis integration was used 
on the intervals which contain singularities; in the remainder
intervals the standard Gaussian rule was used. 
A few iterations were sufficient to achieve a convergence to 
accuracy $10^{-5}$ on all points of the energy mesh.

Below, we present the results obtained for zero-temperature
low-density neutron matter relevant to the physics of neutron star
crusts. The density is parametrized in terms of the Fermi momentum 
$\rho = p_F^3/(3\pi^2)$ and we consider the range $0.4 \le p_F\le 0.6$
fm $^{-1}$ which is below  the maximum of the on-shell pairing gap at about
$p_F\simeq 1$ fm$^{-1}$. The gap decreases at larger densities due to the
repulsive component of the pairing force and we need to include (at
least) $\rho$-meson-exchange to reproduce this feature.
We supplement our input parameters with a set of dimensionless
(i.e., normalized by the density of states) Landau 
parameters. We keep them constant throughout the range of
Fermi momenta considered. The effective Landau mass 
is set to $m^*/m = 0.8$ and we assume $g' = 0.6$, although
we discuss its variations towards the end of this section. To evaluate
the BCS contribution (\ref{DBCS}) we choose  
the parameter values $f_0 = -0.3$ and $g_0 = 0.3$ which are taken from  
Ref. \cite{WAMBACH1}. These values of Landau parameters satisfy the 
well-know stability criteria.

\begin{figure}[t] % fig 
\begin{center}   
{\psfig{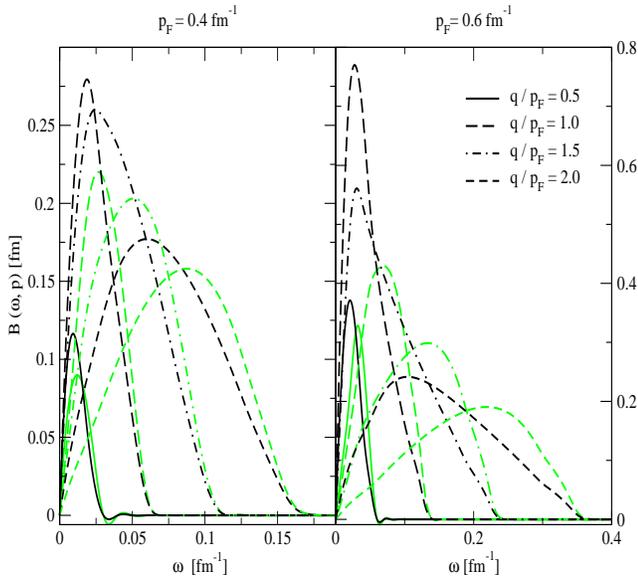}}         
\end{center}
\caption{
The frequency dependence of the free (light lines) and RPA
renormalized (heavy lines) pion spectral function at
several fixed values of the momentum transfer $q/p_F = 0.5$,  1.0, 1.5
and 2.0. The density corresponds to $p_F = 0.4$ fm$^{-1}$ in the 
left panel and to $p_F = 0.6$ fm$^{-1}$ in the right panel.
 }
\label{MSfig:fig2}
\end{figure} 
Figure 2 displays the frequency dependence of the spectral function of
pions at several fixed momentum transfers for densities 
corresponding to $p_F = 0.4 $ (left panel) and 0.6 fm$^{-1}$ (right panel).
It can be seen that the contribution of processes 
with  momentum transfers $q \le p_F/2$  to the spectral function is unimportant
compared to the contribution from large momentum transfers $q \ge  p_F$. 
At fixed $q$ the spectral function has a Lorentzian shape as a function of
energy transfer. For very large momentum 
transfers $q\le  2 p_F $, as the particle-hole excitations with 
higher energies become accessible, the spectral function is  
broadened and its maximum is shifted towards larger energies. 
Compared to the free case, the maximum of the spectral functions 
computed within the RPA is shifted to lower energies while the 
width remains nearly constant.

\begin{figure}[t] % fig 
\begin{center}            
{\psfig{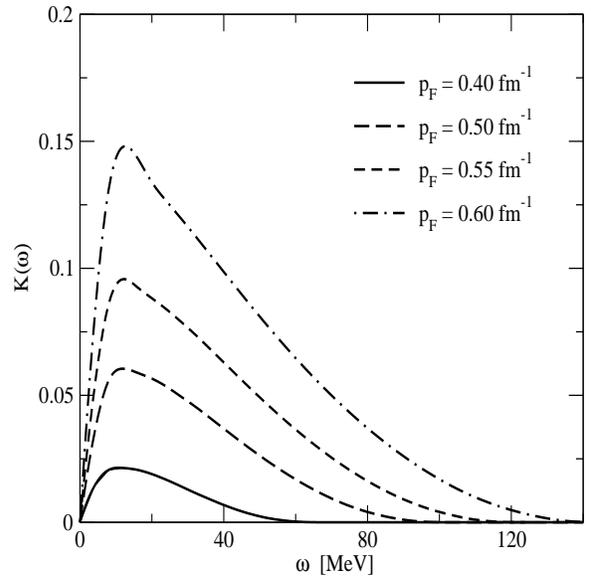}}
\end{center}
\caption{ 
The frequency dependence of the
averaged over the momentum transfer (dimensionless) kernel $K(\omega)$
derived from the RPA renormalized  pion spectral function for 
$p_F = $ 0.4 (solid), 0.5 (long-dashed), 0.55 (dashed) and 0.6
(dashed-dotted lines).
 }
\label{MSfig:fig3}
\end{figure} 
Figure 3 displays the frequency dependence of the 
interaction kernel $K(\omega)$ for several Fermi momenta. 
The shape of the kernel $K(\omega)$ is  Lorentzian  at 
low and intermediate energies [characteristic also to spectral
function $B(\bq,\omega)$ for fixed values of $\vert q\vert$]
with an excess contribution in the high-energy tail. 
The contribution from small momentum transfers
$ q \ll p_F $ to the  kernel $K(\omega)$ 
is negligible,  since the spectral function for these processes 
is generally small  (see Fig. 2) and, in addition, the integrand in Eq. (\ref{K})
is weighted by the factor $q^3$, with each pion-nucleon 
vertex contributing a power  of $q$. The latter factor cuts off
the contribution from small momentum transfers.
The main contribution to the peak of $K(\omega)$ function originates from
the momentum transfers $\vert q \vert \sim p_F $; in this regime 
the maximum of the spectral function 
coincides with that of the $K(\omega)$ function. The excess contribution  
to the high-energy tail of the $K(\omega)$ function is due to the
excitations with large momentum transfers and the magnifying effect
of the pion-nucleon vertices (a factor $\propto q^2$). 
As the density is increased, the $K(\omega)$ function extends to
higher energies, its maximum increase by roughly an order of magnitude over the 
Fermi momentum range $0.4 \le p_F\le 0.6$ fm$^{-1}$ with a minor shift of its 
position to the higher energies. 
Note that the observed increase in the interaction
strength will saturate at densities corresponding to $p_F\le 1$ fm$^{-1}$
once the repulsive interaction via $\rho$ meson-exchange 
will become important.

\begin{figure}[t] % fig 
\begin{center}            
{\psfig{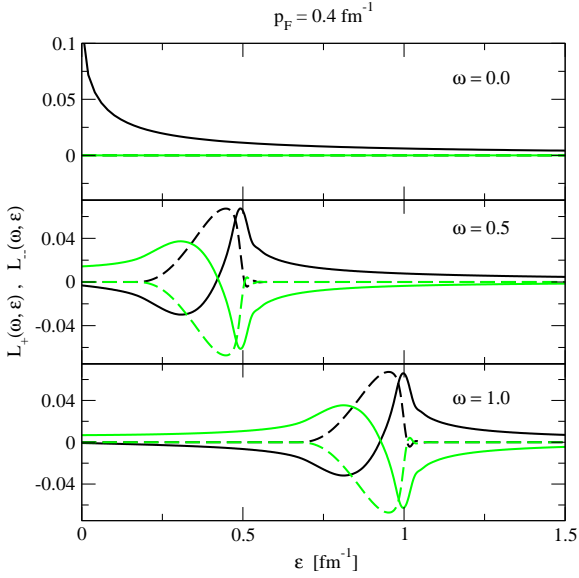}}
\end{center}
\caption{
The $\ep$-dependence of the effective interaction 
kernels $L_+(\omega,\ep)$ (heavy lines) and $L_-(\omega,\ep)$ (light
lines)  for fixed values of $\omega = 0$, 0.5, and 1 fm$^{-1}$ 
for $p_F = 0.4$ fm$^{-1}$. 
The solid and dashed lines refer to the real and imaginary parts. 
 }
\label{MSfig:fig4}
\end{figure} 
The panels in Fig. 4 show the energy dependence of the 
real and imaginary parts of the interaction kernels
$L_{\pm}(\omega,\ep)$ [Eq. (\ref{32bis})] at several fixed values  of the 
integration variable $\ep$. Separating the real and imaginary parts 
in Eq. (\ref{32bis}) it is easy to see that (i) the real parts 
acquire their maximum when $\ep = \omega$  and (ii) 
the imaginary parts vanish for $\ep\ge\omega$ 
features that are evident in Fig. 4. Thus, the main contribution to
the kernel of the gap equation comes from the `diagonal' part
of the interaction kernel $L_{\pm}(\ep,\ep)$.

The frequency dependence of the real and imaginary parts of the
gap function and the wave-function renormalization is shown in Fig. 5. 
The gap function has been replaced by a new function $\Delta'(\omega) =
Z^{-1}(\omega) \Delta(\omega)$ (the prime is dropped below), which 
is equivalent to a redefinition of the energy scale $\omega' = 
\omega Z(\omega)$. The $\omega \to 0$ limit  of these functions
is consistent with what is known from the on-shell physics. The 
real part of the gap function $\Delta_1(\omega = 0)$ is of the 
order of MeV and increases with density, as is the case for 
the pairing gaps deduced using potential models.
The real part of the wave-function renormalization
$Z_1(\omega = 0)$ is {\it larger} than unity, i.e., there is an
enhancement of the density of states at the Fermi surface. 
The imaginary parts of these functions vanish in the
limit $\omega\to 0$ (as they should).
Off the mass-shell the real and imaginary parts of the 
gap function develop a complicated structure: in the low-energy  domain,
below the maximum of the function $K(\omega)$ at about 10 MeV, the real
part of the gap function dominates the imaginary part:
$\vert \Delta_1(\omega)\vert > \vert
\Delta_2(\omega)\vert$. Beyond the maximum the opposite is true 
up to the energies $\omega \simeq 100$ MeV.
\begin{figure}[b] % fig 
\begin{center}            
{\psfig{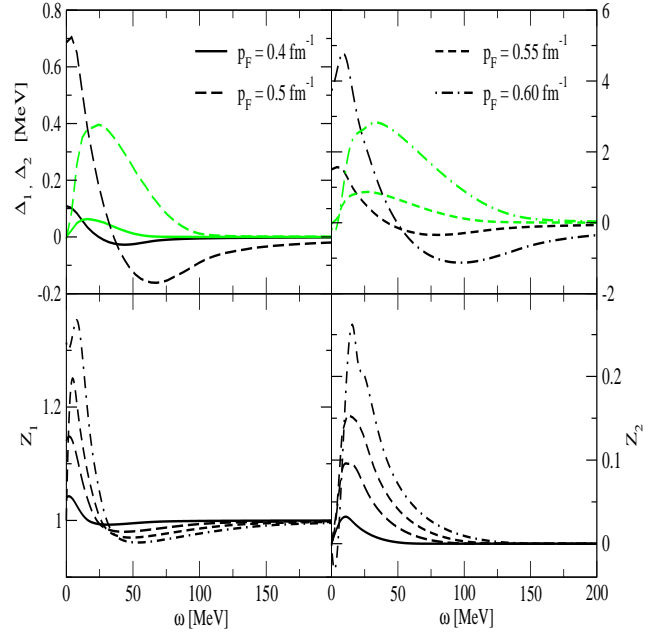}}
\end{center}
\caption{
Top panel: The frequency dependence of the real (heavy lines) 
and imaginary (light lines) parts of the  gap function
$\Delta_1(\omega)$   and  $\Delta_2(\omega)$ for 
$p_F=0.4$ (solid) 0.5 (long-dashed) 0.55
(dashed) and 0.6 (dashed-dotted lines) fm$^{-1}$. 
The on-shell values of 
the pairing gap are 0.1, 0.7, 1.4, and 3.7 MeV for 
$p_F=0.4$, 0.5, 0.55 and 0.6 fm$^{-1}$.
Bottom panel: The frequency dependence of the real (left panel) 
and imaginary (right panel) parts 
of the wave-function renormalization $Z_1(\omega)$  and  
$Z_2(\omega)$. The labeling is the same as in the top panel.
}
\label{MSfig:fig5}
\end{figure} 
A nonvanishing imaginary part $\Delta_2(\omega)$  implies
a finite decay time for the pairing correlations due to the coupling
of the neutron quasiparticles to the pionic modes.   
The imaginary part of the wave-function 
renormalization $Z_2(\omega)$  implies, likewise, finite
lifetime  for the normal quasiparticle excitation.
It is worthwhile to note that the latter effect 
is {\it not} independent of the pairing correlations, since it emerges from
a self-consistent solution of the coupled dynamics of the normal and 
anomalous sectors. The enhancement of the gap and the
wave-function renormalization  with increasing density can be 
traced back to the enhancement of the attractive interaction kernel 
$K(\omega)$. Including the repulsive components of the interaction by
treating heavier (notably $\rho$ meson) exchanges would lead to a
saturation of the attractive interactions at about $p_F\simeq 1$
fm$^{-1}$ and a depression of the gap at larger densities.

Many-body calculations of the Landau parameters lead to values $g'\simeq 0.6-0.8$
\cite{Dickhoff:qr,Dickhoff:qr2,Dickhoff:qr3}. 
The experimental evidence from the studies of spin-isospin modes in 
finite nuclei suggests that 
$g'\simeq 0.6$. 
\begin{figure}[t] % fig 
\begin{center}            
{\psfig{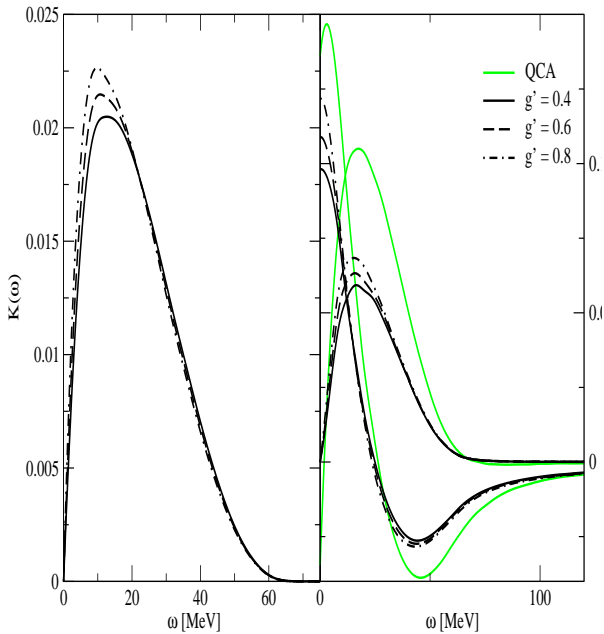}}
\end{center}
\vskip 1cm
\caption{
The frequency dependence of the effective interaction kernel
$K(\omega)$ (left graph) and the real $\Delta_1(\omega)$ and imaginary
$\Delta_2(\omega)$ parts of the pairing gap (right graph) 
for three values of the Landau parameter $g' = 0.4$, 
0.6 and 0.8 and  $p_F = 0.4$ fm$^{-1}$.
The light lines correspond to the QCA approximation with $g' = 0.6$
for $p_F = 0.4$ fm$^{-1}$.
 }
\label{MSfig:fig6}
\end{figure} 
Figure 6 displays the changes in 
the interaction kernel $K(\omega)$ and the pairing 
gap function $\Delta(\omega)$ as the Landau parameter
$g'$ is varied in the range 0.4-0.8. The larger $g'$ the stronger is
the attractive interaction kernel and, hence, the larger is the 
pairing gap. The
overall range of variations of these quantities is, however, small and 
the observables are insensitive to the variations of $g'$.
As discussed in the Appendix, the ordinary QCA
requires that the characteristic off-shell energies 
of the excitation be smaller than the Fermi energy $\omega/\mu\ll 1$ -
a condition which is not well satisfied for excitation energies
in the present model. One finds typically $\omega/\mu\sim 1$. 
The IQCA relaxes the $\omega/\mu\ll 1$ assumption of the QCA, 
but treats the pairing gaps and the wave-function 
renormalizations independent of the momentum variable.
Figure 6 (right panel) compares the results obtained within 
the QCA and the IQCA. Although the qualitative picture is the same in
both approximations, the IQCA leads to smaller gap functions (and
wave-function renormalizations) because the integrations 
over the on-mass-shell energy are limited to a finite shell, 
while the  QCA  assumes infinite integration limits. Calculations which include 
full momentum dependence of the gap function and wave-function 
renormalization will be needed to gauge the remaining error due 
to the IQCA.
\begin{figure}[t] % fig 
\begin{center}            
{\psfig{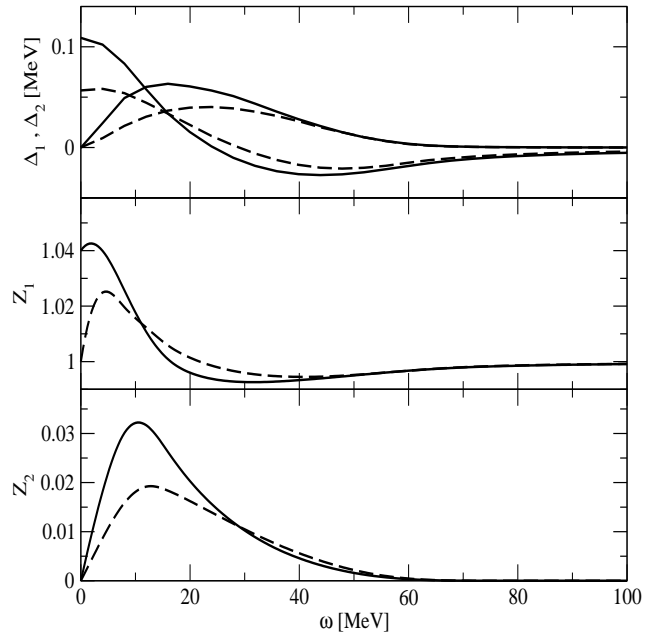}}
\end{center}
\vskip 1cm
\caption{
The frequency dependence of  the real $\Delta_1(\omega)$ and imaginary
$\Delta_2(\omega)$ parts of the pairing gap (top panel) and the real 
$Z_1(\omega)$ and imaginary $Z_2(\omega)$ parts of wave-function 
renormalization (bottom panels) for $p_F = 0.4$ fm$^{-1}$. 
The solid and dashed curves correspond to the results obtained 
with the RPA renormalized and free pion spectral function.
 }
\label{MSfig:fig7}
\end{figure} 
Finally, Fig. 7 compares the results obtained with the free 
and the RPA renormalized pion spectral function. The shapes of the 
gap and renormalizations functions are qualitatively similar in 
both cases; however, the RPA renormalization enhances the values 
of the gap at small frequencies and, therefore, their on-shell 
values. More generally, according to Fig. 7, the RPA
renormalization enhances (suppresses) the gap and renormalization 
functions in the low-frequency (high-frequency) limit. These features
are consistent with the frequency dependence of the free and renormalized 
spectral functions displayed in Fig.~2.

\section{Concluding remarks}

In this work we proposed a model of pairing in nuclear
systems based on the meson-exchange picture of nuclear interactions
that includes retardation effects. The model is illustrated on a
simplest possible example - neutrons interacting via exchange of
neutral pions; the short-range correlations are included 
in the spirit of Landau-Migdal theory of Fermi liquids. Unlike the
treatments based on the potential models, the modifications of the
meson properties in the medium are taken into account by relating the
pairing gap function and the wave-function renormalization to the
spectral functions of mesons. We computed these medium
modifications for the pionic modes within the  non-perturbative 
particle-hole (RPA) resummation scheme, which softens their 
spectrum, but has no qualitative effect on the retardation. 
The model builds in the retardation effect via the 
coupled Fock-exchange self-energies for normal and anomalous sectors. 
The Hartree self-energies receive contributions only
from the short-range part of the interaction, as pions do not
contribute. 
Although we obtain the exact self-energies for our system, these are
solved numerically with the improved quasiclassical approximation (IQCA). 
Since it is difficult to estimate {\it a priori} the error introduced 
through the IQCA (which fixes the momentum transfer at the
Fermi momentum) our numerical results should be taken with caution.

The numerical calculations above show that the real and imaginary
parts of the gap function are of the same order of magnitude 
over a wide energy range. At low
energies, the real part of the gap function dominates the imaginary
part, i.e., the Fourier transform of the anomalous pair-correlation 
function in the time-domain shows a damped oscillatory behavior; 
the picture is opposite in the high-energy regime, 
typically above tens of MeV, where the oscillations are damped out
within a `cycle'.
Physically, the complex pairing gap implies a finite lifetime of the
Cooper pairs due to the emission/absorption of pions. The
real part of the wave-function renormalization (derived
self-consistently with the gap equation) is larger than unity,
i.e., implies an enhancement of the density of states; the finite
imaginary part describes the damping of the normal quasiparticle
excitations due to the coupling to the pion modes. 
The frequency dependence of the pairing gap and the wave-function 
renormalization will have implications on
the responses of nuclear systems to those electromagnetic and weak 
probes which are sensitive to the energy range where these functions
vary substantially.

Extrapolating from the analysis of this paper, it is clear that the
meson-exchange picture of pairing with retardation
can be extended to treat problems which  have been
difficult to handle within the potential models. Such an extension could
incorporate the dynamics of nucleon resonances, for example, to
clarify the role of delta's in the effective retarded interaction
or to deduce their pairing properties. Another interesting direction 
is the application of the  model to the strange sector to study 
the pairing properties of the strange baryons that are stable 
in the interiors of compact stars (typically $\Sigma$ and $\Lambda$ hyperons, 
see Ref. \cite{GLENDENNING,WEBER}), the 
role of the kaon dynamics on the pairing properties of the strange 
baryons, etc. The model appears also suitable for  treating 
the fermionic pair condensation of the baryons on the same 
footing with the Bose condensation of pions or kaons.

\section*{Acknowledgments}
This work has been supported by a grant provided by 
the SFB 382 of the DFG. Helpful conversations with 
Herbert M\"uther and Peter Schuck are gratefully acknowledged.

\section*{Appendix}

The purpose of this appendix is to discuss in detail the
approximations to the full normal and anomalous Green's functions 
that allow us to remove the explicit momentum dependence of the 
wave-function renormalization and the gap function.
We start by examining the momentum dependence of 
Eqs. (\ref{24})-(\ref{26}) in the low-temperature limit.
Since the formal structure of the normal and anomalous 
self-energies is the same, we shall apply our arguments only to the 
normal self-energy, Eq. (\ref{24}).  Below, we assume isotropic medium 
with a spherical Fermi surface.  

\begin{widetext}
Using the spectral representations of the Green's functions 
(\ref{spec1}) and (\ref{spec2}), Eq. (\ref{24}) can be transformed 
to the following form 
\be\label{a1}
&&\Sigma^R(\omega, \bp) = \int\frac{d^3q}{(2\pi)^3} 
{\rm Tr}\{\Gamma_0(\bq) \Gamma(\bq)\}\int_0^{\infty}\frac{d\omega'}{2\pi}
B(\omega',\bq)
\Biggl[g(\omega')G^R(\bp-\bq, \omega+\omega')
\nonumber\\&&
-g(\omega')G^R(\bp-\bq, \omega-\omega')
+\int_{-\infty}^{\infty} \frac{d\ep}{\pi}J(\ep, \omega , \omega')
{\rm Im} G^R(\bp-\bq,\epsilon) \Biggl],  
\ee
where the function $J(\ep, \omega , \omega')$  is defined by 
Eq. (\ref{DEF_J}). Directing the axis of a spherical system of 
coordinates along the vector $\bp$, the integration measure becomes
$d^3q = q^2dq~dx ~d\phi$, where $x$ is the 
cosine of the angle between the vectors $\bp$
and $\bq$. Upon defining $\bk = \bp-\bq$ and replacing the integration
over $x$ by an integration over $\xi_k$,  Eq. (\ref{a1}) becomes
\be\label{a2}
&&\Sigma^R(\omega, p_F)=\frac{m^*}{(2\pi)^2p_F}
\int_0^{\infty}\!\! q dq 
{\rm Tr}\{\Gamma_0(\bq) \Gamma(\bq)\}\int_0^{\infty}
\frac{d\omega'}{2\pi}B(\omega',\bq) H(\omega,\omega'),
\ee
with
\be\label{a3}
H(\omega,\omega')&=&
g(\omega')\int_{\xi_-(q)}^{\xi_+(q)}\!\! d\xi_k~
[G^R(\bk, \omega+\omega')+G^R(\bk, \omega-\omega')]\nonumber\\
&+& \int_{-\infty}^{\infty} \frac{d\ep}{\pi}
J(\ep, \omega , \omega')
\int_{\xi_-(q)}^{\xi_+(q)}\!\! d\xi_k~{\rm Im} G^R(\bk,\epsilon) ,
\ee
where $\xi_k = k^2/2m^*-\mu$, the integration limits are
$\xi_{\pm} = (p_F\pm q)^2/2m^* - \mu$, $m^*$ is the
effective mass, $\mu$ is the chemical potential. Without 
loss of generality,  the momentum argument of the self-energy 
$\Sigma^R(\omega, p)$ is taken as the Fermi momentum, i.e., 
the expressions (\ref{a2}) and (\ref{a3}) are still exact.

{\it Quasiclassical Green's functions}. Now we turn to the
$\xi_k$-integrations in Eq. (\ref{a3}). 
Formally, the quasiclassical approximation (QCA) amounts to 
(i) taking infinite limits $\xi_{\pm}\to \pm \infty$ and (ii)
approximating the function $\Delta(\xi_k,\omega)$ 
and $Z(\xi_k,\omega)$ 
in the propagators by their values on the Fermi surface
$\Delta(\xi(p_F),\omega)$ and $Z(\xi(p_F),\omega)$.
The integrals over the normal and anomalous
Green's functions are then computed by a contour integration
\be\label{tmp1}
&&\int_{-\infty}^{\infty}\!\! d\xi_k~
\frac{\omega Z(\omega)+\xi_k}{\omega^2 Z(\omega)^2-\xi_k^{2}
-\Delta^{R}(\omega)^2+i~{\rm sgn}(\omega)\eta}\nonumber\\
&&\hspace{2cm} =  
-i\pi\frac{\omega Z(\omega) {\rm sgn}(\omega)}{
\sqrt{\omega^2Z^2(\omega)-\Delta(\omega)^2}
}~\theta[\omega^2Z^2(\omega)-\Delta(\omega)^2]
,\\
\label{tmp2}
&&\int_{-\infty}^{\infty}\!\! d\xi_k~
\frac{-\Delta^{R}(\omega)}{\omega^2 Z(\omega)^2-\xi_k^{2}
-\Delta^{R}(\omega)^2+i~{\rm sgn}(\omega)\eta}\nonumber\\
&&\hspace{2cm} = i\pi\frac{\Delta(\omega) {\rm sgn}(\omega)}{
\sqrt{\omega^2Z^2(\omega)-\Delta(\omega)^2}}
~\theta[\omega^2Z^2(\omega)-\Delta(\omega)^2].
\ee
The second approximation [$\xi(k)\to \xi(k_F)$], assumes that 
the gap function and the wave-function renormalization are 
smooth functions of momentum, therefore, at low temperatures,
they can be approximated by their values at the Fermi surface.
For the range of possible momentum transfers $0\le q\le 2p_F$ for 
particles on their Fermi surface,
the functions need to be constant in the interval $k\in [0~;~3p_F]$.
Such an assumption is not inconsistent with the results of the
potential models, but an  error estimate will depend on the
shape of the chosen interaction in the momentum space. Note that 
in our case, the momentum dependence of the functions  is   
controlled by the shape of the spectral function which contributes
mainly for $q\ge p_F$. 

The first approximation (infinite integration limits)
uses the fact that the large momentum
transfers  are important in the integration in 
Eq. (\ref{a2}), so that the limits $\vert \xi_{\pm} \vert \sim \mu $ 
are large compared to
values of the poles of the Green's functions in 
Eqs. (\ref{tmp1}) and (\ref{tmp2}), i.e., the condition
$\xi_{\rm pole} \ll \vert \xi_{\pm}\vert$ is satisfied. 
The main contribution to the $q$ integration comes 
from large momentum transfers $p_F\le q \le 2p_F$ (see Sec. III), so 
that $\xi_{\pm} = (p_F\pm q)^2/2m^* - \mu \sim \pm\mu$ 
[note that {max}~$(\xi_+) = 8\mu$ and  {min}~$(\xi_-) = -\mu$].
In addition, the pion-nucleon vertices contribute a factor 
of $q^2$ to the integrand in Eq. (\ref{a2}), which 
cuts off the contributions from small momentum transfers. The poles
of the Green's functions in Eqs. (\ref{tmp1}) and (\ref{tmp2}) 
are defined by the condition $\vert \xi_{\rm pole} \vert = 
\sqrt{\omega^2Z(\omega)^2-\Delta(\omega)^2}\le \omega$. 
The QCA is then applicable whenever the condition
$\xi_{\rm pole}/\mu\ll 1$ is satisfied.
In our problem, the typical energy scale for the off-shell 
excitations is of the order of  20 MeV which is 
of the same order of magnitude as the Fermi energies;
the parameter $\xi_{\rm pole}/\mu$ could be of the order of unity,
in which case the conventional QCA will break down.

{\it Improved  quasiclassical Green's functions}. 
Here we shall improve on the second approximation of the QCA by 
keeping the integration limits finite. To
decouple the on-shell energy integrations over the Green's functions  
from the remainder of the kernel, we approximate the 
$q$-dependent integration limits  $\xi_{\pm}(q)$ by constants 
$\xi_{\pm}(p_F)$.  Setting the momentum transfer to $q=p_F$
is motivated by the shape of the pion-spectral function 
which, as noted above, contributes essentially in the range 
$p_F\le q\le 2p_F$. 

The integrations in Eq. (\ref{a3}) and the analog equation for the 
anomalous self-energy require then two types of integrals ($\xi
\equiv \xi_k$ hereafter)
\be 
I_1 &=& \int_{\xi_-}^{\xi_+}\!\! 
\frac{d\xi}{\omega^2 Z(\omega)^2-\xi^{2}
-\Delta^{R}(\omega)^2+i~{\rm sgn}(\omega)\eta},\\
I_2 &=& \int_{\xi_-}^{\xi_+}\!\! \frac{\xi~d\xi}{\omega^2 Z(\omega)^2-\xi^{2}
-\Delta^{R}(\omega)^2+i~{\rm sgn}(\omega)\eta}.
\ee
Upon defining a shorthand $D = \omega^2
Z(\omega)^2-\Delta^{R}(\omega)^2$, the first integral can be rewritten
as 
\be 
I_1 &=& \int_{\xi_-}^{\xi_+}\!\!d\xi_k \left[
\frac{\theta(D)}{D-\xi^{2}+i~{\rm sgn}(\omega)\eta}
+\frac{\theta(-D)}{D-\xi^{2}+i~{\rm sgn}(\omega)\eta}\right].
\ee
The first term contains a pole and we apply 
the Dirac identity $1/x = P/x-i\pi\delta(x)$ to separate the real and
imaginary parts ($P$ stands for the principal value). 
The second term is regular and we take the limit $\eta\to 0$
to obtain
\be
I_1 &=&\int_{\xi_-}^{\xi_+}\!\!d\xi
\Biggl\{\theta(D)\frac{ P}{D-\xi^{2}}
-\frac{\theta(-D)}{-D+\xi^{2}}
-i\,\pi\, {\rm sgn}(\omega)\frac{\theta(D)}{2\vert \xi\vert}
\left[\delta(\xi-\sqrt{D})+\delta(\xi+\sqrt{D})\right]\Biggr\}.
\nonumber\\
\ee
These integrals are standard (e.g. 3.3.21 and 3.3.23 of
Ref. \cite{ABRAM}) and the final result is 
\be\label{I1}
I_1 &=& \frac{\theta(D)}{2\sqrt{D}} ~{\rm ln}\Bigg\vert
\frac{(\sqrt{D}+\xi_+)}{(\sqrt{D}-\xi_+)}
\frac{(\sqrt{D}+\xi_-)}{(\sqrt{D}-\xi_-)}
\Bigg\vert + \frac{\theta(-D)}{\sqrt{D}}{\rm arctanh}
\left[\frac{\xi_++\xi_-}{\sqrt{D}(1+\xi_+\xi_-)}\right]
\nonumber\\
&-&i\,\pi\, {\rm sgn}(\omega)\frac{\theta(D)}{2\sqrt{D}}
\left\{
\theta[(\xi_+-\sqrt{D})(\sqrt{D}-\xi_-)]
+\theta[(\xi_++\sqrt{D})(-\sqrt{D}-\xi_-)]
\right\}.\nonumber\\
\ee
\end{widetext}
The step functions in the last term insure that the poles of the
propagators are counted only if they lie within 
the integration limits; to obtain the final expression we have used
the relation arctan($ix) = i~ $arctanh$(x)$.
In the limit $\xi_{\pm}\to\pm \infty $ we recover the right-hand 
sides of Eqs. (\ref{tmp1}) and (\ref{tmp2}).
The integral $I_2$ is computed as above
and, upon using 3.3.22 of Ref. \cite{ABRAM}, we obtain
\be\label{I2}
I_2 = \frac{1}{2}~ {\rm ln}\Bigg\vert
\frac{D-\xi_-^2}{D-\xi_+^2}\Bigg\vert
\left[\theta(D)-\theta(-D)\right].
\ee 
Note that this integral vanishes when the limits are infinite 
($\xi_{\pm}\to\pm \infty $) or, more generally, 
when the integration is within finite limits which are 
symmetrical ($\vert\xi_-\vert = \vert \xi_+ \vert)$.
Under the time reversal the integrals $I_1$ and
$I_2$ transform according to 
\be
{\rm Re}~ I_{1,2}(-\omega) = {\rm Re}~ I_{1,2}(\omega), \quad 
{\rm Im}~ I_1(-\omega) = -{\rm Im}~ I_1(\omega).
\ee
These transformation properties were used to derive Eqs. 
(\ref{31bis})-(\ref{DBCS2}).

%\newpage

\end{document}